\PassOptionsToPackage{table}{xcolor}
\documentclass[conference]{IEEEtran}
\usepackage[utf8]{inputenc}  %
\usepackage{amsmath,amssymb,amsfonts}
\usepackage{algorithmic}
\usepackage{graphicx}
\usepackage{textcomp}
\usepackage{comment}
\usepackage{xspace}
\usepackage{makecell}
\usepackage{adjustbox}
\usepackage{paralist}
\usepackage[shortlabels,inline]{enumitem}
\usepackage{booktabs}
\usepackage[hidelinks]{hyperref}
\usepackage{circledsteps}
\usepackage{balance}

\usepackage{multirow}
\usepackage{color, colortbl}
\definecolor{lightgray}{gray}{0.9}
\definecolor{Gray}{gray}{0.9}

\usepackage[framemethod=tikz]{mdframed}
\mdfdefinestyle{mystyle}{%
  rightline=true,
  innerleftmargin=5,
  innerrightmargin=5,
  outerlinewidth=1pt,
  topline=false,
  leftline=true,
  bottomline=false,
  skipabove=\topsep,
  skipbelow=\topsep,
  backgroundcolor = lightgray,
}

\usepackage[backend=biber, abbreviate=true, dateabbrev=true, alldates=year, isbn=false, doi=false, urldate=comp, url=false, maxbibnames=1, backref=false, style=numeric-comp, language=american, giveninits=true, sortcites=true, sorting=none]{biblatex}
\AtEveryBibitem{\clearlist{location}} %
\AtEveryBibitem{\clearfield{series}} %
\AtEveryBibitem{\clearlist{language}} %

\usepackage{citation-cleaner}

\RemoveBibField{lamprecht2020}{editor}
\RemoveBibField{strubell2019:energy}{publisher}
\RemoveBibField{georgiou2022:green}{publisher}

\newcommand{\puball}{494\xspace} %
\newcommand{\pubrel}{293\xspace} %
\newcommand{\pubSURVEY}{202\xspace} %
\newcommand{\pubrelSURVEY}{108\xspace} %

\newcommand{\pubavail}{79\xspace} %

\newcommand{\cnum}[1]{%
\CircledParamOpts{inner color=white, outer color=black, fill color=black}{1}{\footnotesize\textsf{#1}}}

\newcommand{\COtwo}{\ensuremath{\textrm{CO}_2}}

\newcommand{\contrib}{\par\noindent\textbf{Contributions:}\space}
\newcommand{\head}[1]{\par\noindent\textbf{#1:}\space}

\makeatletter
\def\ps@IEEEtitlepagestyle{%
  \def\@oddfoot{\mycopyrightnotice}%
  \def\@evenfoot{}%
}
\def\mycopyrightnotice{%
  \hspace*{3mm}\includegraphics[width=2cm]{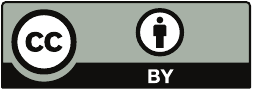}%
  \hspace*{2mm}\raisebox{2.5mm}{%
          \parbox{\columnwidth}{\footnotesize This work is licensed under a Creative Commons \\ Attribution 4.0 International (CC BY 4.0) license.}%
          \hspace*{-68pt}\mbox{1}\hspace{12pt}\fbox{\parbox{.94\columnwidth}{\footnotesize\textsl{Accepted for publication in the 17th ACM/IEEE International Symposium on Empirical Software Engineering and Measurement (ESEM 2023).}}}%
  }%
  \gdef\mycopyrightnotice{}%
}
\makeatother

\begin{document}

\title{An Exploratory Literature Study on Sharing and Energy Use of Language Models for Source Code}

\author{\IEEEauthorblockN{Max Hort}
\IEEEauthorblockA{\textit{Simula Research Laboratory}\\
Oslo, Norway \\
maxh@simula.no}
\and
\IEEEauthorblockN{Anastasiia Grishina}
\IEEEauthorblockA{\textit{Simula Research Laboratory} \& \\
\textit{University of Oslo} \\
Oslo, Norway \\
anastasiia@simula.no}
\and
\IEEEauthorblockN{Leon Moonen}
\IEEEauthorblockA{\textit{Simula Research Laboratory} \& \\
\textit{BI Norwegian Business School} \\
Oslo, Norway \\
leon.moonen@computer.org}
}

\maketitle

\begin{abstract}
\textsc{Context:}
Large language models trained on source code can support a variety of software development tasks, such as code recommendation and program repair.
Large amounts of data for training such models benefit the models' performance. 
However, the size of the data and models results in long training times and high energy consumption.
While publishing source code allows for replicability, users need to repeat the expensive training process if models are not shared.

\textsc{Goals:} The main goal of the study is to investigate if publications that trained 
language models for software engineering (SE) tasks share source code and trained artifacts. The second goal is to analyze the transparency on training energy usage.

\textsc{Methods:} We perform a snowballing-based literature search to find publications on language models for source code, and analyze their reusability from a sustainability standpoint.

\textsc{Results:}
From a total of \puball unique publications, we identified \pubrel relevant publications that use language models to address code-related tasks. 
Among them, 27\% (\pubavail out of \pubrel) make artifacts available for reuse. 
This can be in the form of tools or IDE plugins designed for specific tasks or task-agnostic models that can be fine-tuned for a variety of downstream tasks. %
Moreover, we collect insights on the hardware used for model training, as well as training time, which together determine the energy consumption of the development process.

\textsc{Conclusion:} We find that there are deficiencies in the sharing of information and artifacts for current studies on source code models for software engineering tasks, 
with 40\% of the surveyed papers not sharing source code or trained artifacts. 
We recommend the sharing of source code as well as trained artifacts, to enable sustainable reproducibility.
Moreover, comprehensive information on training times and hardware configurations should be shared for transparency on a model's carbon footprint.

\end{abstract}

\begin{IEEEkeywords}
sustainability,
reuse,
replication,
energy,
DL4SE.
\end{IEEEkeywords}

\section{Introduction}
\noindent
The FAIR data principles are designed to support and enhance the reusability of digital research objects following four guiding principles: to be findable, accessible, interoperable, and reusable~\cite{wilkinson2016:fair}.
While the initial focus of FAIR was on scientific data, the principles have been transferred to research software~\cite{lamprecht2020:fair}. 
Publishing source code supports the replicability of software but may incur repeated training costs, if a software product is data-driven.
Training costs can be especially high for tools that are trained on large amounts of data, such as Machine Learning (ML) models, which have achieved state-of-the-art performance in various disciplines (e.g., text and image understanding, video content prediction~\cite{sun2019:videobert,brown2020:language}).
In particular, Deep Learning (DL) often achieves performance improvements by increasing the amount of training data and the size of the model, 
leading to long training times and substantial energy consumption~\cite{strubell2019:energy}, with an increase in computational costs for state-of-the-art models by a factor of 300\,000 between 2012 and 2018~\cite{schwartz2020:green,georgiou2022:green}.
This trend not only raises barriers for researchers with limited computational resources~\cite{tipirneni2022:structcoder}, it is also harmful to the environment~\cite{schwartz2020:green,strubell2019:energy}.

One class of DL models that benefit from training on large amounts of data are Large Language Models (LLMs). 
LLMs have been able to learn semantic information via training on texts from the Internet and achieve high performance on Natural Language Processing (NLP) tasks~\cite{mikolov2013:distributed,strubell2019:energy}.
Similarly, by training language models on a large corpus of source code (e.g., as provided by GitHub\footnote{~\url{www.github.com}}), one can learn semantic information of source code~\cite{chen2021:evaluating}
and apply the models on SE tasks, such as code generation, bug prediction and fixing, to alleviate developers from tedious work~\cite{lu2021:codexglue}.
This research area is referred to as DL4SE, and the models are referred to as \emph{Source Code Models} (SCMs).

Training an SCM can take more than 100 days and incur high costs from hardware and energy requirements~\cite{zhou2020:hulk,lescao2022:bloom}.
From an energy usage
point of view, only sharing the source code to train the model is wasteful, because replication or reuse requires repeating the expensive and energy-consuming training process. 
Instead, trained models should be considered \emph{digital artifacts} that must be shared to lower the bar for building on existing work~\cite{katz2021:working}.
For instance, fine-tuning an existing task-agnostic model requires only a fraction of the computational costs of training such a model from scratch~\cite{zhou2020:hulk}.

Despite the benefits of sharing the trained models and source code, a large number of studies in DL, including many in DL4SE, do not make code or models publicly available. 
Liu et al.~\cite{liu2021:reproducibility} surveyed deep learning studies in SE conferences and journals. 
They found that 74.2\% of the studies did not share the source code and data for replication. 
Failing to share the data or trained artifacts contradicts the software sustainability-quality characteristics~\cite{condori-fernandez2019:software}. 
Software sustainability is defined from economic, environmental, social and technical dimensions in that software should generate economic value, enable equal and sustained access to social resources, minimize harm to the environment, and ensure technical improvements and maintainability~\cite{lago2015:framing}. 
In this study, we focus on the technical and environmental aspects of sustainability in software, namely reusability and efficiency~\cite{condori-fernandez2018:software,condori-fernandez2019:software}.
To investigate the reusability and resource efficiency of source code models, we perform an exploratory literature search of existing DL4SE publications.
For each publication, we investigate whether code and trained models are available, and what the training and energy requirements are.
In other words, we focus on the following two research questions:
\begin{enumerate}[label=\textbf{{RQ\arabic*:}}, ref=RQ\arabic*, left=0pt]
\item How many DL4SE publications share source code and/or trained models or related trained artifacts? 
\item How much energy was used to train these models? 
\end{enumerate} 

\medskip
\contrib{} The contributions of this paper include:
\begin{compactitem}\renewcommand{\labelitemi}{$\star$}
\item We conduct an exploratory study on the sustainability and reusability of (large) language models for source code. We analyze to what extent publications make trained artifacts available, so that software developers and researchers can reuse and profit from large models trained with high energy consumption without incurring such training costs themselves.
\item We investigate the information provided in \pubavail publications with shared artifacts;
\item We estimate the energy needed for training models from 30 publications that provided sufficient information;
\item We summarize the lessons learned while studying the academic literature with this focus on sustainability;
\item We provide recommendations to help researchers make their models more sustainable and support clearly communicating the relevant aspects in their publications. 
\end{compactitem}

\section{Related Work}
\label{section:rw}

\subsection{Sustainable Software Engineering}

\noindent
Sustainable software engineering addresses sustainability in two regards: (1) creating software that empowers sustainable applications and (2) creating software in a sustainable resource-efficient way.
The former is referred to as Information Technology for Green (IT for Green)~\cite{tomlinson2010:greening} or sustainability \emph{BY} software~\cite{condori-fernandez2019:software}. 
The latter is called Green IT~\cite{dick2010:enhancing} or sustainability \emph{IN} software~\cite{condori-fernandez2019:software}. 
In this study, we focus on sustainability \emph{IN} software and use the term \emph{sustainable software} or \emph{sustainable software engineering} to define reusable shared software that is built with resource usage considerations in mind. %
Development of sustainable software can be supported by integrating sustainability goals in the development process~\cite{dick2010:enhancing}. 
One way to improve the sustainability of software is to optimize its performance by refactoring the source code, which can have positive impacts on the accompanying energy consumption~\cite{hindle2015:green}.
For example, Verdecchia et al.~\cite{verdecchia2018:empirical} showed that refactoring code smells in Java applications can reduce energy consumption by almost  50\%.

In addition to observing the energy consumed by applying software and potential positive effects by providing sustainable solutions, the energy consumed during the development process is of relevance as well, as pointed out by the GREENSOFT model~\cite{naumann2011:greensoft}. %
The GREENSOFT model presents a life cycle for software products.
Accordingly, a green software product should be sustainable during the course of the life cycle, including the software engineering process and the tasks developers address during implementation and maintenance.
To alleviate their workload, they can use tools to automate and support software engineering tasks.
In this regard, Martinez et al.~\cite{martinez2022:energy} addressed the field of green software research by measuring energy consumption induced by development and maintenance activities, in particular Automated Program Repair (APR). 
APR is used to fix software bugs,  which usually incur a high monetary cost to resolve, without requiring manual intervention of developers.
While APR tools tend to report their performance in terms of number of bugs they are able to fix, Martinez et al.~\cite{martinez2022:energy} considered their energy consumption as an additional quality measure. 
To evaluate the trade-off between accuracy and energy consumption of APR tools, they computed the energy cost for each point of accuracy (i.e., energy consumption divided by accuracy).

\subsection{Energy Consumption of Machine Learning Models}
\noindent
The energy consumption of training and developing ML models is becoming a growing concern~\cite{verdecchia2022:datacentric}, with models requiring large amounts of computational resources to train, causing financial costs and \COtwo{} emissions~\cite{georgiou2022:green,garcia-martin2019:estimation}.
Recently, implementation challenges and leaderboards have been introduced to incentivize the development of energy efficient models~\cite{li2016:evaluating,henderson2022:systematic}.
Another proposition is to measure the performance of ML models not only with regard to accuracy, but also to consider energy consumption and trade-offs between the two metrics.

\begin{figure*}[t]
    \centering
    \includegraphics[width=0.999\linewidth, trim={0mm 4mm 0mm 0mm}]{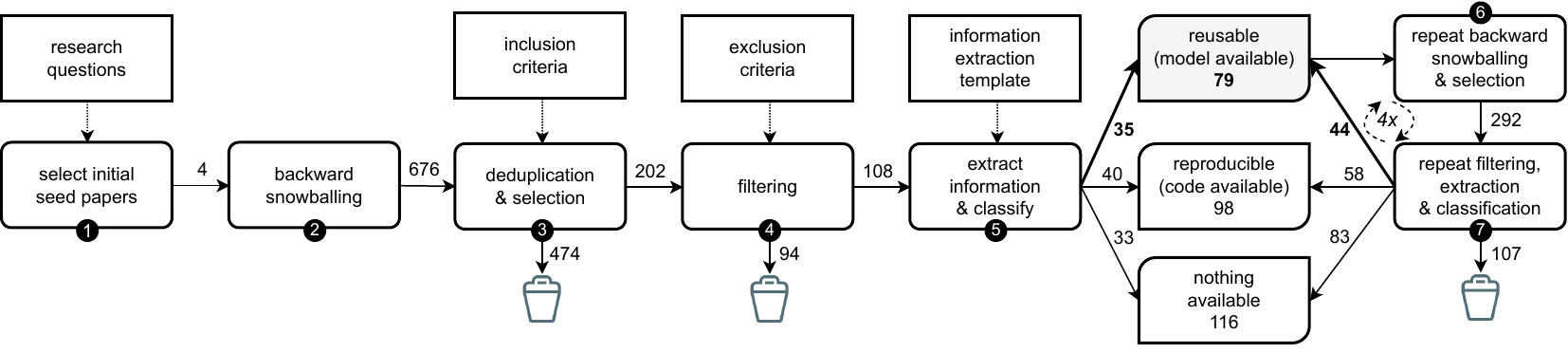}    %
  \caption{Overview of the search procedure.}
  \label{figure:search} 
\end{figure*}

To account for the sustainability--accuracy trade-off, Guti{\'e}rrez et al.~\cite{gutierrez2022:analysing} analyzed the impact of changing solvers for ML models. 
Having applied the models to credit card fraud data, they found configurations that required 2.9x more energy while improving accuracy by only 0.016.
This illustrates that developers can make trade-offs between energy consumption and ML quality measures, such as precision and recall. 
In the same line of research, Georgiou et al.~\cite{georgiou2022:green} compared the energy consumption of two frameworks (\textsc{TensorFlow}, \textsc{PyTorch}) for the development of DL by implementing and comparing the performance of six machine learning models. 
Energy consumption varied significantly in both the training and inference stages, with \textsc{TensorFlow} requiring less energy for training and \textsc{PyTorch} less energy for inference.
However, the framework documentation did not provide information on hardware specifications to allow developers to select models and frameworks with regard to energy requirements.

Verdecchia et al.~\cite{verdecchia2022:datacentric} modified the underlying datasets for training DL models to reduce energy consumption during training.
Results showed that reducing dataset size, either in the number of features or number of data points, improves the energy efficiency by up to 92\%, while having a negligible effect on accuracy reduction for selected algorithms.
Garcia-Martin et al.~\cite{garcia-martin2017:identification} investigated the impact of parameter tuning on energy consumption and accuracy for the Very Fast Decision Tree algorithm.
In some cases, small reductions in accuracy ($< 0.1$) can reduce energy consumption by more than 70\%.
For an overview of publications addressing Green AI (AI systems developed with sustainability and costs considered), we refer to the systematic review by Verdecchia et al.~\cite{verdecchia2023:systematic}.

\subsection{Energy Consumption of Large Language Models}

\noindent
To support responsible NLP, Zhou et al.~\cite{zhou2020:hulk} proposed the platform \textsc{Hulk} for benchmarking pre-trained language models in terms of time and cost. 
Processing time and costs are measured according to cloud services' hardware specifications and resource consumption.
The cost of NLP models is evaluated at three stages: pre-training, fine-tuning, and inference.
Pre-training is the most expensive stage in the development of language models: it can take several days and can cost up to 75,000\$.
However, once pre-trained, a model can be fine-tuned for several tasks, which requires less computational resources.

Strubell et al.~\cite{strubell2019:energy} provided insights on the financial and environmental costs of training large language models for NLP tasks. 
In particular, they estimate the training cost in USD and carbon emissions for four open-source models. %
For example, training large language models with neural architecture search can cause \COtwo{} emissions 17 times as high as the average per-capita consumption in America.
Given the high cost of NLP models, Strubell et al. formulated three actionable recommendations: 
(1)~authors should report training times to allow for a cost-benefit analysis rather to solely focus on accuracy; 
(2)~researchers need equal access to computational resources; 
(3)~efficient hardware and algorithms should be prioritized.

\begin{table*}[t]
\caption{Description of software engineering tasks. Type illustrates the input and output of the respective tasks\\(NL = Natural Language, PL = Programming Language, V = extracted or predicted value).
}
\centering
\label{table:tasks}
\begin{adjustbox}{max width=\textwidth}
\rowcolors{1}{}{lightgray}
\begin{tabular}{lll}
\toprule
Task & Type & Description \\ \midrule
Inference & PL $\rightarrow$ V & Predict code properties (e.g., variable names, data types, code authors). \\
Code summarization & PL$\rightarrow$ NL & Summarize source code snippets in natural language. \\
Code search & NL$\rightarrow$ PL & Search for code snippets given a description. \\
Code completion & PL$\rightarrow$ PL & Recommend likely tokens for a code sequence. \\
Default detection & PL$\rightarrow$ V & Determine whether a code snippet is faulty. \\
Documentation generation & PL$\rightarrow$ NL & Generate (e.g., comments, commit messages, docstrings) \\
Code generation & NL$\rightarrow$ PL & Generate source code given a description (e.g., log messages, program synthesis). \\
Comprehension & \makecell[l]{PL x NL $\rightarrow$ V  \\ PL x NL $\rightarrow$ NL } & \makecell[l]{Check if a code snippet can answer a question.  \\ Answer question about a code snippet.} \\
Clone detection & PL$\rightarrow$ V &  Clone and similarity determination of two code snippets. \\
Code translation & PL $\rightarrow$ PL & Translate source code from one programming language to another. \\
Code repair & PL $\rightarrow$ PL & Repair a faulty code snippet. \\ \bottomrule
\end{tabular}
\end{adjustbox}
\end{table*}

\section{Literature Search}
\label{section:search}

\noindent
To find relevant literature, we adopt and adapt a snowballing search procedure~\cite{jalali2012:systematic, wohlin2014:guidelines, petersen2015:guidelines}. 
We make small adjustments to the search procedure described by Wohlin et al.~\cite{wohlin2014:guidelines}, as we aim to build on four recent surveys in the domain of deep learning models for software engineering tasks. 
The surveys examine different research questions than ours but consider the same domain, which makes them good starting points.
Moreover, we apply the inclusion and exclusion criteria after each snowballing step to control the scope of the search, as it can quickly become too wide and cover all of SE.
Figure~\ref{figure:search} presents an overview of the search procedure and the number of publications collected.
The search and subsequent information extraction were conducted by the first two authors; the third author helped mitigate classification discrepancies where needed.
In step~\cnum{1}, we select the four survey papers that seed the study. We include both published work and arXiv preprints to ensure timeliness. The four seed surveys are:
\begin{itemize}
    \item Chen and Monperrus~\cite{chen2019:literature}: A literature study on embeddings learned from source code.
    Embeddings have been trained on different levels of granularity (e.g., binary code, tokens, functions). A list of 21 publicly available embeddings is provided.
    \item Sharma et al.~\cite{sharma2021:survey}: A survey of ML techniques for analysing source code. A total of 364 studies published from 2002--2021, divided over 12 SE tasks. For each task, data collection, feature extraction and model training stages are outlined.
    They listed 61 tools for analyzing source code and applying ML techniques.
    \item Watson et al.~\cite{watson2022:systematic}: A literature review of deep learning approaches in SE research. A total of 128 deep learning publications spanning 23 SE tasks have been reviewed.
    \item Niu et al.~\cite{niu2022:deep}: A survey on pre-trained models on source code applied to SE tasks. They presented a total of 20 pre-trained code models that have been applied to 18 tasks.
\end{itemize}
These four initial studies contain references to a total of 676 publications (step~\cnum{2}).
After deduplication, we consider a total of \pubSURVEY unique publications for further investigation based on their title  (step~\cnum{3}).
We deem a paper of interest for further analysis if the title matches the following inclusion criteria: 
\begin{enumerate}[label=\textit{\textbf{IC-\arabic*:}}, ref=IC-\arabic*, left=0pt]
\item the publication addresses an SE task, and
\item the publication applies a deep learning technique.
\end{enumerate}
To filter relevant publications, we read all \pubSURVEY publications, published from 2012-2022, 
and flag publications for exclusion if they did not train language models for source code, 
i.e., the exclusion criterion for step~\cnum{4} is:
\begin{enumerate}[label=\textit{\textbf{EC-\arabic*:}}, ref=EC-\arabic*, left=0pt]
\item the publication does not train a source code model.
\end{enumerate}
This step leaves us with \pubrelSURVEY publications for further analysis.

In step~\cnum{5}, we extract information from the \pubrelSURVEY publications to determine how they share artifacts.
First, we investigated if source code is available.
For this purpose, we analyzed the respective publications for links or references to external sources (e.g., a GitHub repository for source code, or Zenodo for datasets and tools).
Among the \pubrelSURVEY publications, 33 publications did not provide source code (``unavailable'' in Fig.~\ref{figure:search}).\footnote{~To ensure that no artifacts were overlooked, we performed additional Google searches for publications that did not mention artifacts for replication.}
Next, we determined if the shared artifacts do not only provide source code, 
but include fully functional tools or checkpoints for ML models that are ready to use, without the need to be trained.
This was the case for 35 out of the \pubrelSURVEY publications (``reusable'' in Fig.~\ref{figure:search}).
The remaining 40 publications provided source code but no trained artifacts (``reproducible'' in Fig.~\ref{figure:search}).

The initial survey-based search is followed by repeated backward snowballing in steps~\cnum{6} \& \cnum{7}.
During snowballing, we collect additional relevant publications that have been cited by the 35 publications which provided trained artifacts collected prior.
Snowballing is performed incrementally on publications that share trained artifacts, 
until no new publications are found (i.e., we perform multiple iterations, 
and stop when a fixed point is reached, which happened after four iterations).
This yielded 292 additional publications (published from 2002-2023) that fit the inclusion criteria, 
bringing the total to \puball (202 from step \cnum{3} and 292 from step \cnum{6}).
After further inspection, 107 of the 292 additional publications did not train a language model and were excluded. 
Furthermore, 83 of those publications did not provide source code, 
and 58 of the publications shared source code but not the trained artifacts.
Thus, repeated snowballing adds 44 publications that share trained artifacts, 
bringing the total to \pubavail publications with shared artifacts, published from 2015 to 2022.

\begin{figure}[t]
  \centering
  \includegraphics[width=0.9\linewidth,trim=0 5mm 0 2mm]{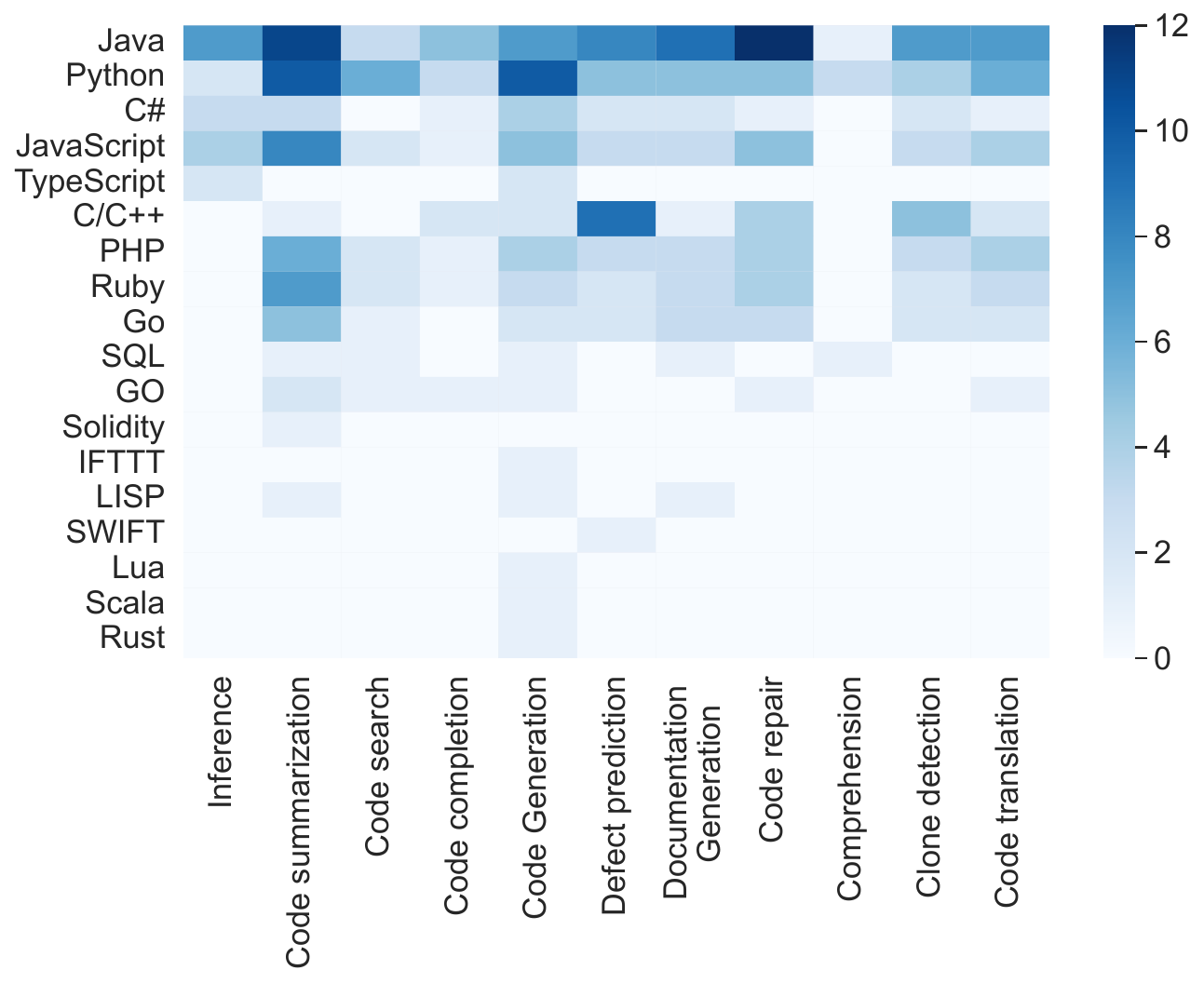}
  \caption{Number of publications with shared artifacts for combinations of task and programming language.}
  \label{fig:heatmap} 
\end{figure}

We classify these \pubavail publications with respect to the 11 SE tasks presented in Table~\ref{table:tasks}, which were inspired by Niu et al.~\cite{niu2022:deep}.
To address these tasks, source code models were trained on 18 different programming languages.
The most frequent languages include Java (45 publications), Python (32 publications), and C and/or C++ (18 publications). 
Figure~\ref{fig:heatmap} presents the number of publications for each combination of programming language and SE task (e.g., an approach trained on Java source code for code completion).

We found that there are two types of trained artifacts that were publicly available: 
(1) trained ML models and tools; (2) source code embeddings.
While trained models and tools are aimed at a specific task, source code embeddings are task-agnostic and provide comprehensive code representations for training future models with less effort than generating pre-trained embeddings~\cite{zhou2020:hulk}.
Section~\ref{section:specific} (task-specific tools) and Section~\ref{section:agnostic} (task-agnostic embeddings) present detailed information for the two types of shared artifacts.

\begin{mdframed}[style=mystyle]
\head{Answer to RQ1} Out of the reviewed \pubrel publications, 33\% shared source code and 27\% shared trained artifacts.
\end{mdframed}

\begin{table*}[t]
\caption{Training details for task-specific language models. For each model, we list hardware details, training time in $hours$ and estimated energy consumption in $kWh$, if the information is available. If only partially trained artifacts are available, we append the approach reference with \textsuperscript{\textdagger}.
}
\centering
\label{table:tools}
\newlength{\firstcol}
\setlength{\firstcol}{6.5em}
\setlength{\tabcolsep}{3pt}
\begin{adjustbox}{max width=\textwidth}
\rowcolors{1}{}{lightgray}
\begin{tabular}{lllll@{}rr}
\toprule
Task & Approach & Year & Language & Hardware & Hours & kWh \\ \midrule
\cellcolor{white}& DeepSim~\cite{zhao2018:deepsim}\textsuperscript{\textdagger} & 2018 & Java & desktop PC: Intel i7 CPU at 4.0GHz, 4 cores; 1 NVIDIA RTX 1080 GPU & 4 & 1.1 \\
\cellcolor{white}& FCDetector~\cite{fang2020:functional}\textsuperscript{\textdagger} & 2020 & C++ & server: 2.4GHz CPU, 8 cores; 1 NVIDIA GTX 1080 GPU & 2 & 1.1 \\
\cellcolor{white}& func2vec~\cite{defreez2018:pathbased}\textsuperscript{\textdagger} & 2018 & C & AWS, Intel Xeon E5-2686 v4 (Broadwell) Processors and DDR4 Memory. & 2 &  \\
 \multirow{-4}{\firstcol}{\cellcolor{white}Clone detection} & TBCCD~\cite{yu2019:neural}\textsuperscript{\textdagger} & 2019 & Java, C &  &  &  \\ \midrule
\cellcolor{white}& C3PO~\cite{brody2020:structural} & 2020 & C\# & 1 NVIDIA Tesla V100 GPU & 9 & 4.6 \\
\cellcolor{white}& AnyCodeGen~\cite{alon2020:structural} & 2020 & Java, C++ & NVIDIA Tesla V100 GPU &  &  \\
\cellcolor{white}& Retrieve-and-edit~\cite{hashimoto2018:retrieveandedit}\textsuperscript{\textdagger} & 2018 & Python &  &  &  \\
 \multirow{-4}{\firstcol}{\cellcolor{white}Code completion} & Codit~\cite{chakraborty2022:codit} & 2022 & Java &  &  &  \\ \midrule
 \cellcolor{white}& Intellicode compose~\cite{svyatkovskiy2020:intellicode} & 2020 & Python, C\#, JavaScript, TypeScript &  80 NVIDIA Tesla v100 GPUs, 32 GB & 492.5 & 19,750.0 \\
\cellcolor{white}& Pythia~\cite{svyatkovskiy2019:pythia} & 2019 & Python & NVIDIA Tesla V100 GPUs & 30,000 & 15,330.0 \\
\cellcolor{white}& Bayou~\cite{murali2018:neural} & 2018 & Java & AWS, p2.xlarge: 1 NVIDIA K80 GPU, 12 GB  & 10 & 4 \\
\cellcolor{white}& Xu et al.~\cite{xu2020:incorporating} & 2020 & Python &   &  &  \\
 \multirow{-5}{\firstcol}{\cellcolor{white}Code generation} & NL2code~\cite{yin2017:syntactic} & 2017 & Python, IFTTT &  &  &  \\ \midrule
\cellcolor{white}& CURE~\cite{jiang2021:cure} & 2021 & Java & server: 1  NVIDIA TITAN V GPU, 3 NVIDIA TITAN Xp GPUs & 528 & 448.8 \\ 
\cellcolor{white}& RLAssist~\cite{gupta2019:deep} & 2019 & C & Intel Xeon E5-2630 v4 CPU at 2.20GHz, 32 GB & 504 & 81.6 \\
\cellcolor{white}& TFix~\cite{berabi2021:tfix} & 2021 & JavaScript & 8 NVIDIA RTX 2080 Ti GPUs & 96 & 321.6 \\
\cellcolor{white}& Sequencer~\cite{chen2019:sequencer} & 2019 & Java & 1 NVIDIA K80 GPU & 1 & 0.6 \\
\cellcolor{white}& DeepRepair~\cite{white2019:sorting} & 2019 & Java &  &  &  \\
\cellcolor{white}& CC2Vec~\cite{hoang2020:cc2vec} & 2020 & Java, C &  &  &  \\
\cellcolor{white}& DL4PatchCorrectness~\cite{tian2020:evaluating} & 2020 & Java &  &  &  \\
\cellcolor{white}& RewardRepair~\cite{ye2022:neural} & 2022 & Java &  &  &  \\
\cellcolor{white}& VRepair~\cite{chen2022:neural} & 2022 & C &  &  &  \\
\multirow{-10}{\firstcol}{\cellcolor{white}Code repair} & Tufano et al.~\cite{tufano2019:learning} & 2019 & Java &  &  &  \\ \midrule
\cellcolor{white}& MP-CAT~\cite{haldar2020:multiperspective} & 2020 & Python &  & 67 &  \\
\cellcolor{white}& CoCLR~\cite{huang2021:cosqa} & 2021 & Python & 1 NVIDIA Tesla V100 GPU, 16 GB  &  &  \\
\cellcolor{white}& CoaCor~\cite{yao2019:coacor} & 2019 & SQL &  &  &  \\
 \multirow{-4}{\firstcol}{\cellcolor{white}Code search}& NCS~\cite{heyman2020:neural} & 2020 & Python &  &  &  \\ \midrule
\cellcolor{white}& CAST~\cite{shi2021:cast} & 2021 & Java & server: 4 NVIDIA Tesla V100 GPUs & 33 & 67.1 \\
\cellcolor{white}& LeClair et al.~\cite{leclair2020:improved} & 2020 & Java & Xeon E1430v4 CPUs, 110 GB; 1 NVIDIA Titan RTX GPU, 1 Quadro P5000 GPU &  &  \\
\cellcolor{white}& Attn-to-FC~\cite{haque2020:improved} & 2020 & Python & Intel Core i7 CPU at 2.2GHz, 64GB 1600 MHz DDR; 1 NVIDIA Titan X GPU, 2GB &  &  \\
\multirow{-4}{\firstcol}{\cellcolor{white}Code summarization} & MMTrans~\cite{yang2021:multimodal} & 2021 & Solidity & server: 4 NVIDIA RTX 2080 Ti GPUs, 11 GB  &  &  \\ \midrule
\cellcolor{white}Code translation & TransCoder~\cite{lachaux2020:unsupervised:arxiv} & 2020 & Java, C++, Python & 32 NVIDIA V100 GPUs &  &  \\ \midrule
\cellcolor{white}& CoCLR~\cite{huang2021:cosqa} & 2021 & Python & 1 NVIDIA Tesla V100 GPU, 16 GB  &  &  \\
\cellcolor{white}& CodeQA~\cite{liu2021:codeqa} & 2021 & Java, Python & 3 NVIDIA 1080 Ti GPUs &  &  \\
\multirow{-3}{\firstcol}{\cellcolor{white}Comprehension} & Staqc~\cite{yao2018:staqc}\textsuperscript{\textdagger} & 2018 & Python, SQL &  &  &  \\ \midrule
\cellcolor{white}& IVDetect~\cite{li2021:vulnerability} & 2021 & C/C++ &  & 239 &  \\
\cellcolor{white}& VulBERTa~\cite{hanif2022:vulberta} & 2022 & C/C++ & \makecell[l]{pre-train: GCP VMs: 48 vCPUs, 240 GB; 2 NVIDIA Tesla A100 GPUs, 40 GB \\ fine-tuning: 48 cores Intel Xeon Silver CPU, 292 GB; 2 NVIDIA TITAN Xp GPUs}  & 106 & 91.0 \\
\cellcolor{white}& DeepWukong~\cite{cheng2021:deepwukong} & 2021 & C/C++ & Intel Xeon E5-1620 CPU at 3.50GHz; 1 NVIDIA RTX 1080 Ti GPU & 23 & 12.7 \\
\cellcolor{white}& Li et al.~\cite{li2019:improving} & 2019 & Java &  & 11 &  \\
\cellcolor{white}& Lin et al.~\cite{lin2018:crossproject}\textsuperscript{\textdagger} & 2018 & C & server: 2 Intel Xeon E5-2690 v3 CPUs at 2.60GHz, 96 GB  & 6 & 3.1 \\
\cellcolor{white}& FUNDED~\cite{wang2021:combining}\textsuperscript{\textdagger} & 2021 & Java, C, C++, PHP, SWIFT & server:  Intel Xeon CPU, 2.4GHz, 14 cores; 1 NVIDIA 2080Ti GPU & 2 & 0.9 \\
\cellcolor{white}& Deepbugs~\cite{pradel2018:deepbugs}\textsuperscript{\textdagger} & 2018 & JavaScript & Intel Xeon E5-2650 CPU, 48 cores, 64 GB; 1 NVIDIA Tesla P100 GPU & 1 & 1.4 \\
\cellcolor{white}& REVEAL~\cite{chakraborty2020:deep:arxiv} & 2020 & C/C++ & 16 Intel Xeon CPUs at 2.60GHz, 252 GB; 1 NVIDIA RTX 1080 Ti GPU &  &  \\
\multirow{-9}{\firstcol}{\cellcolor{white}Defect prediction} & CC2Vec~\cite{hoang2020:cc2vec} & 2020 & Java, C &  &  &  \\ \midrule
\cellcolor{white}& CUP~\cite{liu2020:automating} & 2020 & Java & Intel Xeon CPU at 2.7GHz, 40 cores & 290 & 168.5 \\
\cellcolor{white}& CC2Vec~\cite{hoang2020:cc2vec} & 2020 & Java, C &  &  &  \\
\cellcolor{white}& DEEP JIT~\cite{panthaplackel2020:deep}\textsuperscript{\textdagger} & 2020 & Java &  &  &  \\
\multirow{-4}{\firstcol}{\cellcolor{white}Documentation generation} & DeepCommenter~\cite{li2020:deepcommenter} & 2020 & Java &  &  &  \\ \midrule
\cellcolor{white}& PigeonJS~\cite{alon2018:general}\textsuperscript{\textdagger} & 2018 & JavaScript, Java, Python, C\# &  & 100 &  \\
\cellcolor{white}& Type4Py~\cite{mir2022:type4py} & 2022 & Python & \makecell[l]{ AMD Ryzen Threadripper 1920X with 24 threads at 3.5GHz, 64 GB  \\ 2 NVIDIA RTX 2080 TI GPUs} & 100 & 2.0 \\
\cellcolor{white}& JS NICE~\cite{raychev2015:predicting} & 2015 & JavaScript & 4 Xeon CPUs at 2.13GHz, 32 cores & 10 &  \\
\cellcolor{white}& NL2Type~\cite{malik2019:nl2type} & 2019 & JavaScript & Intel Xeon E5-2650 CPU, 48 cores, 64 GB; 1 NVIDIA Tesla P100 GPU, 16 GB & 4 & 5.7 \\
\cellcolor{white}& DeepTyper~\cite{hellendoorn2018:deep} & 2018 & TypeScript & Intel i7-8700 CPU, 6 cores, 32 GB; 1 NVIDIA RTX 1080 Ti GPU &  &  \\
\cellcolor{white}& Lambdanet~\cite{wei2020:lambdanet} & 2020 & TypeScript &  &  &  \\
\multirow{-7}{\firstcol}{\cellcolor{white}Inference} & NATURALIZE~\cite{allamanis2015:suggesting}\textsuperscript{\textdagger} & 2015 & Java &  &  & \\ \bottomrule
\end{tabular}
\end{adjustbox}
\vspace*{-2ex}
\end{table*}

\section{Task-Specific Code Models}
\label{section:specific}

\noindent
This section presents approaches with shared artifacts that are designed to address specific tasks.
In total, we collected 52 task-specific publications, which are summarized in Table~\ref{table:tools}.
Publications are presented with regards to the task they address and their respective programming language is shown, as well as hardware configuration and training time, if provided.

\begin{table*}[t]
\caption{Training details for task-agnostic language models. For each model, we list hardware details, training time in $hours$ and estimated energy consumption in $kWh$, if the information is available.
}
\centering
\label{table:agnostic}
\setlength{\tabcolsep}{3pt}
\begin{adjustbox}{max width=\textwidth}
\rowcolors{1}{}{lightgray}
\begin{tabular}{p{12em}lp{23em}p{26.5em}@{\hspace{-2em}}rr}
\toprule
Approach & Year & Language & Hardware & Time in hours & kWh \\ \midrule
BLOOM~\cite{lescao2022:bloom} & 2022 & Java, PHP, C++, Python, JavaScript, C\#, Ruby, Lua, TypeScript, GO, C, Scala, Rust & server: 384 NVIDIA A100 GPUs, 80 GB  & 1,082,990 & 433,196 \\
Prophetnet-x~\cite{qi2021:prophetnetx:arxiv} & 2021 & Go, Java, JS, Php, Python, Ruby & NVIDIA Tesla V100 GPUs & 30,000 & 15,330 \\
CodeBERT~\cite{feng2020:codebert:arxiv} & 2020 & Python, Java, JavaScript, PHP, Ruby, Go & server: 16 NVIDIA Tesla V100 GPUs, 32 GB  & 1,320 & 10,610 \\
Dobf~\cite{roziere2021:dobf} & 2021 & Java, Python & 32 NVIDIA V100 GPUs & 192 & 3,080 \\
Codet5~\cite{wang2021:codet5:arxiv} & 2021 & Ruby, JavaScript, Go, Python, Java, Php, C, C\# & server/cluster: 16 NVIDIA A100 GPUs, 40 GB & 288 & 1,930 \\
PLBART~\cite{ahmad2021:unified:arxiv} & 2021 & Java, Python & 8 NVIDIA RTX 2080 Ti GPUs & 276 & 925 \\
Mastropaolo et al.~\cite{mastropaolo2021:studying} & 2021 & Java & Google Cloud, Colab: 8 TPUs, 35.5 GB memory & 343 & 766 \\
Graphcodebert~\cite{guo2021:graphcodebert:arxiv} & 2021 & Ruby, JS, Go, Python, Java, PHP & server: 32 NVIDIA Tesla V100 GPUs, 32 GB  & 83 & 667 \\
CodeTrans~\cite{elnaggar2021:codetrans} & 2021 & Python, Java, JavaScript, PHP, Ruby, Go, C\#, SQL, LISP & 1 TPU v3-8  & 2,088 & 582 \\
GREAT~\cite{hellendoorn2022:global} & 2022 & Python & 1 Tesla P100 GPU & 120 & 51 \\
Javabert~\cite{desousa2021:javabert} & 2021 & Java & 3 NVIDIA Titan X GPUs, 12 GB  & 24 & 30 \\
code2vec~\cite{alon2019:code2vec} & 2019 & Java & 1 NVIDIA Tesla K80 GPU & 36 & 18 \\
OpenVocabCodeNLM~\cite{karampatsis2020:big} & 2020 & Java, Python, C & GPUs & 336 & - \\
GraphCode2Vec~\cite{ma2022:graphcode2vec} & 2022 & Java & server: 40 CPUs@2.20GHz, 256GB; 1 NVIDIA Tesla V100 GPU & - & - \\
Spt-code~\cite{niu2022:sptcode} & 2022 & Java, Python, JavaScript, PHP, GO, Ruby & 4 NIVDIA A100s9 GPUs & - & - \\
StructCoder~\cite{tipirneni2022:structcoder} & 2022 & Java, Python, PHP, JavaScript & 4 RTX 8000 GPUs, 48GB & - & - \\
Codex~\cite{chen2021:evaluating} & 2021 & Python & Azure & - & - \\
Cotext~\cite{phan2021:cotext:arxiv} & 2021 & Python, Java, JavaScript, PHP, Ruby, Go & 1 TPU v2-8 & - & - \\
CuBERT~\cite{kanade2020:learning} & 2020 & Python & TPUs & - & - \\
TSSA~\cite{shrivastava2020:onthefly:cap} & 2020 & Java & 1 NVIDIA P100 GPU, 16 GB; 1 K80 GPU, 16GB memory & - & - \\
CodeGPT~\cite{lu2021:codexglue:arxiv} & 2021 & Python, Java & - & - & - \\
ContraCode~\cite{jain2021:contrastive} & 2021 & JavaScript & - & - & - \\
CodeTransformer~\cite{zugner2021:languageagnostic:arxiv} & 2021 & Python, JavaScript, Ruby, GO & - & - & - \\
DAMP~\cite{yefet2020:adversarial} & 2020 & Java, C\# & - & - & - \\
Obfuscated Code2Vec~\cite{compton2020:embedding} & 2020 & Java & - & - & - \\
code2seq~\cite{alon2019:code2seq} & 2019 & Java, C\# & - & - & - \\
Efstathiou and Spinellis~\cite{efstathiou2019:semantic} & 2019 & Java, Python, C++, C\#, C, PHP & - & - & - \\
\bottomrule
\end{tabular}
\end{adjustbox}
\end{table*}

Among the 52 publications, two publications shared artifacts for more than one task.
Hoang et al.~\cite{hoang2020:cc2vec} proposed CC2Vec, an approach for representing code changes. 
For each of the three tasks (log message generation, bug fixing patch identification, and just-in-time defect prediction), they trained and shared a separate model. 
Huang et al.~\cite{huang2021:cosqa} first introduced a new dataset called CoSQA, consisting of 20,604 human-annotated labels for natural language and source code pairs. Additionally, they proposed a model, CoCLR, trained on two tasks: code search and question answering.
Their GitHub repository provides model checkpoints for both of these tasks.

The most frequently addressed tasks are concerned with faulty programs: code repair and defect prediction. Ten publications proposed approaches for code repair and nine publications addressed defect prediction.
The task with the fewest available artifacts is code translation. Only Lachaux et al.~\cite{lachaux2020:unsupervised:arxiv} shared their TransCoder models for translating between three programming languages (Java, C++, Python).
To allow for the translation of each pair of languages, they shared two models: 1) translate C++ $\rightarrow$ Java, Java $\rightarrow$ C++, Java $\rightarrow$ Python; 2) C++ $\rightarrow$ Python, Python $\rightarrow$ C++, Python $\rightarrow$ Java.

The most popular programming languages, among 11 unique languages considered by the 52 publications, are Java (23 out of 52 publications), C/C++ (14 out of 52 publications), and Python (14 out of 52 publications). 
In detail, 42 publications considered one programming language, while ten publications were applied to more than one language: six publications considered two programming languages, one publication considered three languages, and three publications considered four languages.
This results in an average of 1.33 programming languages considered per publication.

In addition to programming languages considered, we collect training details, such as hardware used and training time for each publication.
However, those are not always provided.
There are 22 out of 52 publications without hardware details (42\%) and 26 out of 52 without training time (50\%), 33\% shared neither information (17 out of 52 publications).
The training time of 26 publications with such details ranges from two hours or less~\cite{chen2019:sequencer,mir2022:type4py,wang2021:combining,defreez2018:pathbased,pradel2018:deepbugs} to hundreds of hours~\cite{svyatkovskiy2020:intellicode,gupta2019:deep}.
While it is common to perform training on GPUs, there are four publications that did not use any GPU for their training procedure, published from 2015--2019~\cite{lin2018:crossproject,gupta2019:deep,defreez2018:pathbased,raychev2015:predicting}.
Commonly, publications used a single GPU for training~\cite{alon2020:structural,fang2020:functional,huang2021:cosqa,hellendoorn2018:deep,cheng2021:deepwukong,chakraborty2020:deep:arxiv,wang2021:combining,malik2019:nl2type,zhao2018:deepsim,chen2019:sequencer,murali2018:neural,pradel2018:deepbugs,haque2020:improved,brody2020:structural}, sometimes in combination with CPUs.
The highest amount of GPUs have been used by Svyatkovskiy et al.~\cite{svyatkovskiy2020:intellicode}. They utilized 5 Lambda V100 boxes, with 16 V100 GPUs each, resulting in 80 GPUs.

While we focus on the training procedure and the energy associated with creating and sharing an ML model, we note the application of such models can vary highly for different SE tasks.
Usually, the reported tested times are lower than the required training time (e.g., more than 100 times quicker than training~\cite{cheng2021:deepwukong,li2019:improving,fang2020:functional}), but in particular, program repair experiments can require long testing times.
For example, Chen et al.~\cite{chen2019:sequencer} applied Sequencer for 130 hours to find patches for 75 bugs. 
White et al.~\cite{white2019:sorting} applied their program repair tool DeepRepair for 2,616 days.
Data extraction and preparation steps can also require considerable amounts of time and compute resources, ranging from 5-12 days~\cite{liu2020:automating,wang2021:combining,li2021:vulnerability}.

The majority of task-specific publications provided access to the full trained models, some of which one needs to request access to~\cite{yin2017:syntactic,li2019:improving}.
Moreover, there are approaches shared as online tools~\cite{murali2018:neural,raychev2015:predicting,alon2020:structural} or IDE extensions~\cite{mir2022:type4py,li2020:deepcommenter,svyatkovskiy2020:intellicode,svyatkovskiy2019:pythia}.
There are also 12 out of 52 publications that did not share the full model, but trained embedding files, which are used by the model. These are marked in Table~\ref{table:tools} with the \textsuperscript{\textdagger} symbol.

\section{Task-Agnostic Code Models}
\label{section:agnostic}

\noindent
This section presents task-agnostic code models which share means of representing source code as embeddings, for a variety of downstream tasks. 
These models are able to transform code snippets to embeddings, which can be fine-tuned to SE tasks.
For example, Lu et al.~\cite{lu2021:codexglue:arxiv} provided fine-tuning details for the CodeXGLUE benchmark, with information for task-specific training and inference time for each task.\footnote{~\url{https://microsoft.github.io/CodeXGLUE/}}
The fine-tuning time ranges from 2 GPU hours (defect detection) to 60 hours (text-to-code generation, documentation translation).

In total, we collected 27 task-agnostic models, as shown in Table~\ref{table:agnostic}.
For each publication, we list the model name and the programming languages it was trained on. If available, we list details on hardware configuration and training times.
Among the 27 publications, 52\% did not provide training time details (14 out of 27) and 26\% did not provide their hardware configurations (7 out of 27). 
For publications without hardware details, training time is not reported as well.

Among the publications that shared training time details, the shortest duration is found for code2vec~\cite{alon2019:code2vec}, which was trained for 1.5 days and a single GPU. 
However, training large models can usually take weeks, up to 87 days for CodeTrans~\cite{elnaggar2021:codetrans} and 3.5 months for BLOOM~\cite{lescao2022:bloom}.
The long training time of BLOOM can be explained by the fact that it was trained on the highest number of programming languages (13 programming languages) in addition to 46 natural languages.
Thereby, BLOOM is also the model trained on the highest number of programming languages, as it was trained on 13 out 14 programming languages we observed. 
BLOOM was not trained on LISP, which was only considered by CodeTrans~\cite{elnaggar2021:codetrans}
On average, each task-agnostic model is trained on source code data from 3.6 programming languages.
Moreover, 10 out of the 27 publications train on a single programming language, which in 6 out 10 cases is Java.

In comparison to task-specific models, task-agnostic models are trained on more programming languages, 3.6 in comparison to 1.3 programming languages on average, and require a higher computational effort.
In addition, publications that provide task-agnostic models for embedding source code are more likely to share hardware configurations than publications with task-specific models.
The proportion of publications without training time details is comparable for both types (50\% and 52\%, for task-specific and task-agnostic models, respectively).
Another difference is that task-agnostic models use more sophisticated hardware for training, with each publication using either GPUs or TPUs. Only one publication considered CPUs in addition to GPUs for training~\cite{ma2022:graphcode2vec}.

\section{Discussion}
\label{section:discussion}

\noindent
To discuss the various facets of RQ2, we consider three aspects: 
(A) How much energy do task-specific and task-agnostic models consume? 
(B) To what extent do studies on source code models take sustainability concerns into account? 
(C) When is sharing a model more efficient than re-training?

\subsection{Energy Usage of
Task-specific vs. Task-agnostic Models}
\label{section:consumed}

\noindent
First, we perform a comparison of the energy consumed by training task-specific and task-agnostic models. 
For this purpose, we collect all publications that provide hardware and training time details, such that we can estimate the consumed energy in kilowatt-hours (kWh). 
In total, 30 publications provide sufficient information.\footnote{Note that we did not contact authors to provide missing information.}

To estimate energy consumption, we used the Green Algorithms calculator~\cite{lannelongue2021:green}.\footnote{~\url{https://www.green-algorithms.org/}}
This calculator is designed to estimate the carbon footprint and energy needed to run algorithms based on the number and type of CPU/GPU cores, runtime, available memory, and platform run on (PC, local server, cloud).
It is also possible to consider the location for training and running algorithms, because the energy mix in the grid impacts the carbon footprint.
In contrast to the \textit{Machine Learning Emissions Calculator}~\cite{lacoste2019:quantifying}, the Green Algorithms calculator provides averaged options when details are missing (e.g., ``world'' if the country is unknown, ``Any'' CPU type if the type is not known), which is beneficial for estimating energy consumption if these details are missing. 
Our estimates report the energy needed in kWh with the default location set to ``world'', because server locations are seldom reported.

We share energy usage estimations in the last column of Table~\ref{table:tools} and Table~\ref{table:agnostic} for task-specific and task-agnostic artifacts, respectively. 
Hardware specifications required by the Green Algorithms calculator are incomplete in the majority of studies considered. 
Most of the models are trained using a type of accelerator, such as GPU or TPU. 
Four studies reported cloud provider utilization, while the other studies used different server configurations.
To this end, we make assumptions about the missing specifications based on the standard CPU and GPU values stated in product descriptions on web pages of Intel and NVIDIA. 
In case the calculator does not cover a specific CPU type, we fetch Thermal Design Power (TDP) information from the manufacturers' website, to estimate the power used per core. 
In addition, for publications that used both CPU and GPU for training their models, 
we consider both to be active during the entirety of the training time, unless stated differently.
We use the specifications reported in Table~\ref{table:specs} unless stated otherwise by the publications.

\begin{table}[t]
\caption{Assumptions about missing hardware specifications made for energy usage estimation with Green Algorithms calculator. 
We specify assumptions made for GPU, CPU and accelerators if hardware information in a publication is incomplete.
} 
\centering
\label{table:specs}
\begin{adjustbox}{max width=\columnwidth}
\rowcolors{1}{}{lightgray}
\renewcommand{\arraystretch}{1.1}
\begin{tabular}{ll}
\toprule
\cellcolor{white}\textbf{Model (NVIDIA)}  & \textbf{GPU Memory Assumptions} \\ 
 \midrule
 Any                & 16 GB, if missing \\
 GTX 1080   & 8 GB   \\
 RTX 2080 Ti        & 16 GB  \\
 Tesla K80          & 12 GB  \\
 Tesla V100         & 16 GB  \\
 Titan P100         & 16 GB  \\
 Titan V            & 12 GB  \\
 Titan Xp           & 12 GB  \\
 \midrule
\cellcolor{white}\textbf{Study}      & \cellcolor{white}\textbf{Assumed Hardware Parameters}   \\ 
 \midrule
 \centering
 CUP~\cite{liu2020:automating} & CPU: Intel Xeon E5-2697 v4, 64 GB \\
 RLAssist~\cite{gupta2019:deep} & CPU: 10 cores; TDP 8.5 W \\
 \cellcolor{white} & \cellcolor{white} \\
 \multirow{-2}{*}{\cellcolor{white}DeepSim~\cite{zhao2018:deepsim}} & \multirow{-2}{15em}{\cellcolor{white}CPU: Intel Core i7-4790K, 32 GB, TDP 22 W} \\
 NL2Type~\cite{malik2019:nl2type}\cellcolor{lightgray}& CPU: TDP 11.875 W\cellcolor{lightgray}\\
 Mastropaolo et al.~\cite{mastropaolo2021:studying} \cellcolor{white} & TPU type v2 \cellcolor{white} \\
 Type4Py~\cite{mir2022:type4py}\cellcolor{lightgray} & CPU: 12 cores, TDP 15 W\cellcolor{lightgray}\\
 \bottomrule    
\end{tabular}
\end{adjustbox}
\end{table}

\begin{figure}[t]
    \centering
      \includegraphics[width=0.8\linewidth,trim={0mm 5mm 0mm 0mm}]{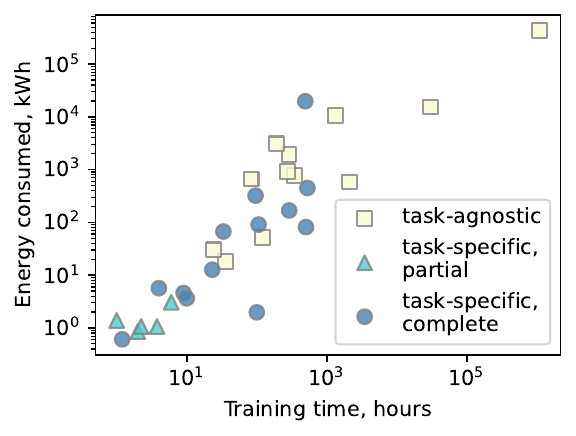}
  \caption{Energy used for training publicly available models for code. We distinguish partially shared and fully shared task-specific models, and fully shared task-agnostic models.}
  \label{fig:energy-use} 
  \vspace*{-2ex}
\end{figure}

Figure~\ref{fig:energy-use} illustrates the energy consumed for training for each of the 30 publications. 
Of these, 12 provided task-agnostic models and 18 task-specific models.
Among the task-specific models, 5 only provided partially trained artifacts (e.g., embeddings that are used for later training), and therefore require additional training effort before usage.

\begin{mdframed}[style=mystyle]
\head{Answer to RQ2-A}
30 out of \pubavail publications share sufficient information to estimate their energy consumption during training. Among these, the training of task-agnostic models used more sophisticated hardware (GPUs and TPUs) and required more energy.
\end{mdframed}

\subsection{Sustainability Concerns Considered in DL4SE Studies}
\label{section:concerns}
\noindent
In Section~\ref{section:consumed}, we outlined publications that provided sufficient information to estimate the energy required to replicate their models (i.e., hardware and training time). 
While this is important to understand how high the energy requirements are, it does not illustrate whether the resource usage is sustainable, or whether sustainability was taken into account.
Only in a few cases do authors consider the sustainability of the training process and the carbon footprint caused.
Here, we present all three publications that, in addition to providing pre-trained artifacts, mention sustainability concerns when training.
All of these three trained and provided large task-agnostic models, two of which required ``hundreds of petaflop/s-days of compute''~\cite{chen2021:evaluating} or more than a million GPU hours for training~\cite{lescao2022:bloom}.

Chen et al.~\cite{chen2021:evaluating} trained Codex on Azure, which purchases carbon credits and uses renewable energies to reduce the carbon footprint.
Using the pre-trained Codex model for repeated inference could exceed training costs.

Wang et al.~\cite{wang2021:codet5:arxiv} stated that the experimental design followed the objective of avoiding unnecessary computation, by creating smaller-sized models in comparison to existing ones, such as Codex.
Moreover, training has been conducted on the Google Cloud Platform, which purchases carbon credits to offset the 49.25kg \COtwo{} caused by training CodeT5.

Le Scao et al.~\cite{lescao2022:bloom} considered various sustainability aspects during the creation of BLOOM: equipment manufacturing, model training, model deployment.
The 81 tons of \COtwo{} needed for training BLOOM can be attributed to 14\% equipment manufacturing, 30\% training, 55\% idle energy consumption.
Training benefits from France's energy grid, which uses nuclear energy in a large proportion, as a low-carbon energy source.
Further details on the carbon footprint of BLOOM are provided in a dedicated study by Luccioni et al.~\cite{luccioni2022:estimating}.

\begin{mdframed}[style=mystyle]
\head{Answer to RQ2-B}
Three publications covered sustainability concerns of the training process in addition to providing trained models. For example, they used cloud providers that purchase carbon credits or calculated \COtwo{} emissions resulting from training the shared models.
\end{mdframed}

\subsection{When is Sharing Models More Efficient than Re-training?}
\label{section:is-it-worth-it}

\noindent
In this section, we provide an exemplary scenario to compute and compare the energy required for training and storing a task-specific and task-agnostic model.
We also show the energy used for downloading shared artifacts, to illustrate the energy-saving capabilities of sharing models trained on code.

\begin{table}[b]
\caption{Comparison of energy consumed for 
training, sharing and downloading pre-trained artifacts for exemplary task-specific and task-agnostic models.
}
\centering
\label{table:compare}
\begin{adjustbox}{max width=\columnwidth}
\begin{tabular}{llrr}
\toprule
                              &                  & Task-specific  & Task-agnostic \\ \midrule
\multirow{3}{*}{Training} & Hardware         &    1 GPU        &     8 GPUs              \\
                              & Training time    &    10 hours         &       250 hours            \\
                              & Memory available &    64 GB       &      64 GB      \\
                              & Platform &    Local Server & Local Server  \\
                              & Location &   World & World     \\
                              & Energy           &    3.74 kWh        &   677.95 kWh                \\ \midrule
\multirow{3}{*}{Sharing}      & Model Size           &   500 MB     &  5 GB               \\
                              & Upload           &   0.003 kWh     & 0.033 kWh                  \\
                              & Storage (1 year) &   0.5 kWh         &  4.95 kWh                 \\ \midrule       

\multirow{3}{*}{Download}      & 100 downloads           &    0.33   kWh    &   3.32 kWh                \\
                              & 1,000 downloads           &    3.32   kWh    &   33.19 kWh                \\
                              & 10,000 downloads &     33.19 kWh       &   331.94 kWh                \\
                              & 1,000,000 downloads         &   3,319.44 kWh         & 33,194.44 kWh \\ \bottomrule    
\end{tabular}
\end{adjustbox}
\end{table}

In accordance with the energy estimates for the training process in Section~\ref{section:consumed}, we used the calculator provided by Lannelongue et al.~\cite{lannelongue2021:green}.
To determine the energy consumption of training and sharing language models, we followed Lakim et al.~\cite{lakim2022:holistic} who provided an assessment of the carbon footprint for the Arabic language model \textit{Noor}.
Data storage energy consumption estimates are based on the cloud storage energy consumption reported by Posani et al.~\cite{posani2019:carbon}, with a mean operating peak power of 11.3 W/TB.
This measure includes a redundancy factor of 2 (i.e., an additional copy is stored) and Power Usage Effectiveness (PUE) of 1.6.
Per year, this results in the energy consumption of 99 kWh per TB of data.
Following the formula by Baliga et al.~\cite{baliga2011:green}, Posani et al.~\cite{posani2019:carbon} estimated the energy consumption of data transfers to be 23.9kJ/GB, with 1kJ being equal to 1/3600 kWh.

In Table~\ref{table:compare}, we illustrate the exemplary energy consumption of sharing a tool (500 MB) and a large task-agnostic model (5 GB) over the span of one year.
Note that we only consider the energy consumed by training and data storage. Other aspects, such as the manufacturing of hardware components, are omitted. Therefore, our example presents a reduced estimate of the complete energy consumption of the entire model lifecycle.

One also needs to consider the rebound effect depending on the number of downloads when estimating potential energy savings~\cite{hilty2011:five}.
If trained models are downloaded because it is easy rather than necessary, then excess downloads can cause higher energy consumption than the initial model training caused.
In our example, this is the case after 1,247 downloads for the task-specific model and 20,544 downloads for the task-agnostic model.
While 20,544 downloads may sound like a large number, CodeBERT~\cite{feng2020:codebert} was downloaded 1,982,300 times from Hugging Face in January 2023.\footnote{~\url{https://huggingface.co/microsoft/codebert-base}}

\begin{mdframed}[style=mystyle]
\head{Answer to RQ2-C}
A rebound effect happens when a shared model is downloaded too many times. For example, energy usage for storage and downloading of a 500~MB-size task-specific model is higher than re-training it after ca.~1,130 downloads.
\end{mdframed}

\section{Threats to Validity}
\label{section:threats}

\noindent
This section discusses the threats to validity of this mapping study based on the categories identified by Zhou et al.~\cite{zhou2016:map}.

\head{Internal Validity}
Internal validity refers to threats to the validity of results presented in this work, for example, due to missing relevant publications during the literature search stage~\cite{martin2017:survey}.
To mitigate this threat, we use a systematic process, starting our literature search with four comprehensive surveys on machine learning approaches for the SE domain. 
These provide an overview of relevant publications from 2022 and prior.
Moreover, we apply four stages of snowballing to find additional references.
This allows us to gather previous approaches which shared their artifacts, but there is a chance that we miss more recent works that have not been cited by any publication in our corpus, as we did not perform forward snowballing. 
While this can slightly alter our results, we are hopeful that 
recent works are more likely to share artifacts than publications from the past 10 years.

\head{External Validity}
External validity addresses the domain to which our findings can be generalized to. 
While our study focuses on the sustainability of shared artifacts for LLMs on code, our results confirm observations of related studies, such as high energy consumption in training LLMs for NLP tasks~\cite{strubell2019:energy} and a lack of shared artifacts of DL studies for SE tasks~\cite{liu2021:reproducibility}.
Therefore, we hope our findings and recommendations are beneficial beyond LLM models for code.

\head{Construct Validity}
Construct validity is concerned with the quality of measures chosen to study the construct of interest.
In our case, we were first interested in whether and which artifacts are shared (i.e., none, source code, trained models). For this purpose, we considered the absolute amount of publications with respect to the amount of shared artifacts, which coincides with the construct we want to measure.

Afterwards, we estimated the energy requirements in kWh for training language models, for which we used the Green Algorithms calculator~\cite{lannelongue2021:green}.
For the validity of estimates, we assume the correctness of the calculator and the information specified in the respective publications (i.e., type of CPU/GPU and training time). 
If the information provided was not sufficient, we had to make choices for available memory and hardware parameters. All such choices are provided in Table~\ref{table:specs}, to make our kWh estimates reproducible.

\head{Conclusion Validity}
Conclusion validity describes whether the operations performed and obtained results in this study (e.g., literature search, data collection) can be reproduced~\cite{zhou2016:map}.
Section~\ref{section:search} outlines our literature search procedure, starting from four existing surveys, followed by iterative snowballing steps.
We list our inclusion and exclusion criteria, to allow for a reproducibility.
Moreover, we provide a link to the collected publications and extracted information in the \textit{Data Availability} section, such that our search results can be verified.
To allow for the reproducibility of observations and results, we provide all the relevant extracted information in Tables~\ref{table:tools} and~\ref{table:agnostic}.
When information is insufficient, we provided all assumptions over hardware specifications in Table~\ref{table:specs}.

\section{Lessons Learned}
\label{section:lessons}

\noindent
\textbf{1.} 
In general, shared information on the amount of energy consumed by the training and use of SCMs is limited (RQ1; some notable exceptions, such as BLOOM~\cite{lescao2022:bloom,luccioni2022:estimating}, RQ2-B).
Specifically, CPU and GPU details are missing or incomplete in the majority of papers, while they are crucial for energy usage estimation.
Even with hardware details and training time available, it is hard to make accurate estimates of energy consumption and \COtwo{} footprint since other missing factors, such as the server location, impact the estimation (RQ2-A).

\noindent
\textbf{2.} From the data that is shared, we see that SCMs are extremely energy-intensive to train due to computational requirements,
in particular, when compared to the energy required for downloading shared artifacts (RQ2-C).
It is therefore important that researchers share their artifacts (RQ1), including pre-trained and fine-tuned models, as well as explore ways to reduce their energy consumption, such as training in clouds with low \COtwo{} emissions (e.g., hydro-powered).

\noindent
\textbf{3.} In general, we find that the larger the model, the higher the energy consumed for its training (RQ2-A), increasing the importance to share model artifacts to ensure sustainability. 

\noindent
\textbf{4.} 
Not only the energy consumption of training but also that of long-term storage of pre-trained models and datasets, as well as of their downloads should be considered (RQ2-C).

\noindent
\textbf{5.} On the positive side, SCMs provide ample opportunities for collaborative and cooperative efforts.
Sharing artifacts in the end can lead to higher sustainability than when all users would develop their models independently. %
More work and data is needed to be able to analyze this trade-off, which is why there is a need for a series of guidelines or a checklist to help people systematically report on the environmental/sustainability impact of their techniques.

\section{Recommendations}
\label{section:recommendations}
\noindent
\textbf{1.} Define the scope of the research and the intended application of task-agnostic or task-specific SCMs to ensure a good understanding of the intended tasks and reuse potential.

\noindent
\textbf{2.} Establish a set of clear and transparent metrics for energy consumption and sustainability to ensure systematic, accurate, and reliable reporting.

\noindent
\textbf{3.} Specify details of the hardware and software configuration used for the training and inference of SCMs, including the exact types of the processors and accelerators, memory and the number of cores for CPU (e.g., Intel i7-8700 CPU, 6 cores, 32GB memory), the model and memory for GPUs (e.g., 1 NVIDIA Titan X GPU, 12GB), as well as storage media and infrastructure (RQ2-A).

\noindent
\textbf{4.} Provide 
energy consumption measurements~\cite{strubell2019:energy,kalaitzoglou2014:practical} or estimations for both training and inference (RQ2-A). Use existing proven calculators~\cite{lacoste2019:quantifying, lannelongue2021:green}
and provide complete details in the paper, not just the final result, so that the computation can be repeated if an improved calculator becomes available.

\noindent
\textbf{5.} Document the \COtwo{} footprint associated with energy consumption, considering energy sources and carbon offsetting applied. For cloud infrastructures, this means including the provider and region, because these details vary by location.

\noindent
\textbf{6.} Assess other environmental impacts of SCMs, including the amount of data and storage required and the impact on the (network) infrastructure (RQ2-C).

\noindent
\textbf{7.} Provide (and promote) open access to data and models to foster collaboration and reduce duplication of efforts, thereby reducing the energy and resource requirements for SCM development and fine-tuning.

Observe that several of these recommendations overlap with the recommendations for reproducible machine learning~\cite{pineau2022:improving}, which also cover additional aspects.

\section{Conclusion}
\label{section:conclusion}

\noindent
In this exploratory study, we have performed a snowballing study (i.e., four iterations of backwards snowballing) to find publications on language models for SE tasks, from which we gathered \puball publications of interest.
After applying our inclusion and exclusion criteria, we are left with \pubrel studies, which we investigated further with regard to their reusability and sustainability (e.g., are trained artifacts shared?).
We showed that there are deficiencies in the existing studies that train language models on source code regarding the transparency of sustainability aspects. %
Among the \pubrel publications, only 27\% provide trained artifacts to enable the reuse of their models without incurring the same amount of training effort; 40\% of the reviewed publications provide neither source code nor trained artifacts.

We collect training information from the surveyed publications, including the hardware configurations and training time. 
This allows us to estimate how much time and resources can be saved by reusing the artifacts or how many resources are needed to replicate the models.
We have estimated the energy consumption for 30 publications that provided sufficient information (i.e., number and type of processors, training time), while only two publications provided details on energy consumption and \COtwo{} of the model training~\cite{wang2021:codet5:arxiv,lescao2022:bloom}.

We stress the importance of describing hardware configurations and processing times, so that even if energy consumption is not reported, one can estimate the required resources and judge whether one wants to spend effort to replicate ML models.
This agrees with Bender et al.~\cite{bender2021:dangers}, who called for the research community to prioritize the environmental and financial cost of deep learning systems, by reporting or evaluating them with regard to resource usage.
Optimally, if a publication creates an ML tool or model with the clear intention of its reuse, it can be beneficial to make trained artifacts available. 
As shown, making small tools available for download and reuse can prevent unnecessary energy consumption as opposed to training tools from scratch.

\head{Future Work}
One possible direction for future investigation is an analysis of the literature that cites the energy calculators~\cite{lacoste2019:quantifying, lannelongue2021:green} mentioned earlier to assess if their use indeed leads to better communication of sustainability aspects. 
This could add further evidence to our recommendations. 

\section*{Data Availability}

\noindent
To support open science and allow for replication and verification of our work, 
an overview of the collected publications and the extracted information
is made available via Zenodo.\footnote{~Replication package on Zenodo: \url{https://doi.org/10.5281/zenodo.8058668}.
}

\section*{Acknowledgements}

\noindent
The research presented in this paper was financially supported by the Research Council of Norway through the secureIT project (grant \#288787). Max Hort is supported through the ERCIM ‘Alain Bensoussan’ Fellowship Programme.

\smallskip

\balance
\printbibliography

 \end{document}